\documentclass[aps,prb,floatfix,showpacs,reprint]{revtex4-1}

\usepackage{graphicx}
\usepackage{dcolumn}
\usepackage{bm}
\usepackage{amsmath}
\usepackage{rotating}
\usepackage{amssymb}
\usepackage{verbatim}
\usepackage{float}

\newcommand{\bra}[1]{\langle #1|}
\newcommand{\ket}[1]{|#1\rangle}

\begin{document}

\title{Comparison of one-dimensional and quasi-one-dimensional Hubbard models from the variational two-electron reduced-density-matrix method}

\author{Nicholas C. Rubin}
\author{David A. Mazziotti}

\email{damazz@uchicago.edu}

\affiliation{Department of Chemistry and The James Franck Institute, The University of Chicago, Chicago, IL 60637 USA }

\date{Submitted March 20, 2014; Revised April 16, 2014}

\begin{abstract}

   Minimizing the energy of an $N$-electron system as a functional of a two-electron reduced density matrix (2-RDM), constrained by necessary $N$-representability conditions (conditions for the 2-RDM to represent an ensemble $N$-electron quantum system), yields a rigorous lower bound to the ground-state energy in contrast to variational wavefunction methods.  We characterize the performance of two sets of approximate constraints, (2,2)-positivity (DQG) and approximate (2,3)-positivity (DQGT) conditions, at capturing correlation in one-dimensional and quasi-one-dimensional (ladder) Hubbard models.  We find that, while both the DQG and DQGT conditions capture both the weak and strong correlation limits, the more stringent DQGT conditions improve the ground-state energies, the natural occupation numbers, the pair correlation function, the effective hopping, and the connected (cumulant) part of the 2-RDM.  We observe that the DQGT conditions are effective at capturing strong electron correlation effects in both one- and quasi-one-dimensional lattices for both half filling and less-than-half filling.

\end{abstract}

\pacs{71.10.-w, 31.10.+z, 71.10.Fd}

\maketitle

\section{Introduction}

The authors are honored to dedicate this article in memory of Isaiah Shavitt whose remarkable contributions transformed electronic structure theory.       Observables that depend on pairwise interactions can be directly computed with the two-electron reduced-density-matrix (2-RDM) without the $N$-electron wave function.\cite{RDM07, CR12, coleman_reduced_Mat}  Integration of the $N$-electron density matrix over all electrons save two yields the 2-RDM
\begin{equation}\label{D2fromDNintro}
^{2}D(12;{\bar 1}{\bar 2}) = \binom{N}{2} \int{^{N}D(12..N;{\bar 1}{\bar 2}..N) d3..dN} .
\end{equation}
Minimization of the energy with respect to the 2-RDM results in unphysical ground states because the variational space of two-electron density matrices is larger than the set of 2-RDMs that can be contracted from an ensemble $N$-electron density matrix.\cite{PhysRev.105.1421}  The constraints on the 2-RDM to ensure a valid $N$-electron density matrix preimage are known as {\em $N$-representability conditions}.\cite{RevModPhys.35.668, JMathPhys.8.2063, Erdahl, coleman_reduced_Mat}  While the set of necessary and sufficient constraints needed to satisfy Eq.~(\ref{D2fromDNintro}) was unknown until recently,\cite{PhysRevLett.108.263002} an approximate class of $N$-representability constraints has been demonstrated to be sufficient for calculating ground-state properties of the metal-to-insulator transition of hydrogen chains,\cite{sinitskiy:014104} ground states and charge distributions of quantum dots,\cite{rothman_variational_2008} quantum phase transitions,\cite{gidofalvi_computation_2006,schwerdtfeger_convex-set_2009} dissociation channels,\cite{PhysRevLett.93.213001} and quantum lattice systems.\cite{hammond_variational_2005, anderson_second-order_2012, verstichel_variational_2012, baumgratz_lower_2012, barthel_solving_2012, Verstichel201312, verstichel_v2dm_2013}

Variational minimization of the energy as a functional of the 2-RDM is expressible as a special convex optimization problem known as a {\em semidefinite program}.\cite{VB96, W97, NN93, PhysRevLett.93.213001, nakata:164113, PhysRevLett.106.083001}  The convexity of the $N$-representable set of 2-RDMs ensures a rigorous lower bound to the ground-state energy. Because the variational 2-RDM method generates large semidefinite programs, the second-order interior-point methods must be exchanged for first-order methods.  We solve the semidefinite program by a boundary-point method developed for RDMs.\cite{PhysRevLett.106.083001} The boundary-point method is one to two orders of magnitude faster than the first-order algorithm, previously applied to the one-dimensional Hubbard model.\cite{hammond_variational_2005}

In this paper we examine the electron correlation of one-dimensional and quasi-one-dimensional Hubbard models with two sets of approximate $N$-representability conditions.  While recent RDM calculations have examined linear\cite{hammond_variational_2005} as well as $4\times4$ and $6\times6$ Hubbard lattices,\cite{anderson_second-order_2012, verstichel_v2dm_2013} there has not been an exploration of RDMs on quasi-one-dimensional Hubbard lattices with a comparison to the one-dimensional Hubbard lattices.  How does the electron correlation change as we move from a one-dimensional to a quasi-one-dimensional Hubbard model?  How are these changes in correlation reflected in the required $N$-representability conditions on the 2-RDM?  One- and two-particle correlation functions are used to compare the electronic structure of the half-filled states of the $1\times10$ and $2\times10$ lattices with periodic boundary conditions.  The degree of correlation captured by approximate $N$-representability conditions is probed by examining the 1-particle occupations around the Fermi surfaces of both lattices and measuring the entanglement with a size-extensive correlation metric, the Frobenius norm squared of the cumulant part of the 2-RDM.\cite{TJ_JCP1252006}

\section{Theory}

In this section we review the salient features of variational reduced density methodology and discuss the approximate $N$-representability conditions used in this work.

\subsection{2-RDM Method}

In variational 2-RDM theory the energy functional is minimized with respect to the 2-RDM
\begin{equation}\label{Efunctional}
E = {\rm Tr}(^{2} K \;^{2}D) ,
\end{equation}
where
\begin{equation}
\label{reducedH}
^{2} K_{kl}^{ij} = \frac{1}{N-1}\left( h^{i}_{k} \delta^{j}_{l} + h^{j}_{l} \delta^{i}_{k} \right) +  u_{ij}^{kl}
\end{equation}
and
\begin{equation}
\label{eq:D22}
^{2} D^{ij}_{kl} = {\rm Tr}( {\hat a}^{\dagger}_{i} {\hat a}^{\dagger}_{j} {\hat a}_{l} {\hat a}_{k}, {}^{2} D ) .
\end{equation}
In Eq.~(\ref{reducedH}) $^{2}K^{ij}_{kl}$ are the elements of the reduced Hamiltonian matrix, in Eq.~(\ref{eq:D22}) $^{2}D^{ij}_{kl}$ are the elements of the 2-RDM in a spin orbital basis set, $h$ and $u$ are tensors containing the one- and two-electron integrals, and the $\hat{a},\left(\hat{a}^{\dagger}\right)$ are the fermionic annihilation (creation) operators.  The variational space in which the energy is minimized can be constrained by a hierarchical set of $N$-representability constraints on the 2-RDM called $(2,p)$-positivity conditions.  These constraints have recently been shown when $p=r$ to form a complete set of $N$-representability conditions\cite{PhysRevLett.108.263002} where $r$ is the rank of the one-electron spin-orbital basis set.

The $(p,p)$-positivity conditions\cite{PhysRevA.63.042113,*M06} on the $p$-RDM restrict the $(p+1)$ metric (or overlap) matrices of the form
\begin{equation}\label{metricOverlap}
M = \langle \psi | \hat{C} \hat{C}^{\dagger} | \psi \rangle
\end{equation}
to be positive semidefinite where the operators $\hat{C}$ are linear combinations of all possible products of $p$ creation and/or annihilation operators.  A matrix $M$ is {\em positive semidefinite}, denoted by $M \succeq 0$, if and only if all of its eigenvalues are nonnegative.  The three distinct (2,2)-positivity conditions\cite{RevModPhys.35.668, garrod:1756, PhysRevA.63.042113, PhysRevA.65.062511, N01, Zhao_T1} are given by
\begin{eqnarray}
^{2} D & \succeq & 0  \label{eq:d2} \\
^{2} Q & \succeq & 0  \label{eq:q2} \\
^{2} G & \succeq & 0 ,\label{eq:g2}
\end{eqnarray}
where
\begin{eqnarray}\label{2Pos}
^{2}D_{ij}^{pq} & = & \bra{\psi}a_{p}^{\dag}a_{q}^{\dag}a_{j}a_{i}\ket{\psi} \\
^{2}Q_{ij}^{pq} & = & \bra{\psi}a_{p}a_{q}a_{j}^{\dag}a_{i}^{\dag}\ket{\psi} \\
^{2}G_{ij}^{pq} & = & \bra{\psi}a_{p}^{\dag}a_{q}a_{j}^{\dag}a_{i}\ket{\psi} .
\end{eqnarray}
Physically, the semidefinite conditions on the $^{2} D$, $^{2} Q$, and $^{2} G$ matrices restrict the probabilities of finding particle-particle, hole-hole, and particle-hole pairs to be nonnegative, respectively.  Even though the nonnegativity constraints in Eqs.~(\ref{eq:d2}-\ref{eq:g2}) are non-redundant, these matrices contain equivalent information as each matrix can be expressed in a one-to-one mapping of another by the fermionic anticommutation relations.  The (2,2)-positivity conditions are often denoted as DQG.  Contraction of the positive semidefinite $^{2} D$, $^{2} Q$, and $^{2} G$ matrices generates 1-particle $^{1} D$ and 1-hole $^{1} Q$ matrices that are also positive semidefinite.

The (2,3)-positivity conditions\cite{PhysRevLett.108.263002} on the 2-RDM may be formed by taking all convex combinations of the (3,3)-positivity conditions that depend only on the 2-RDM.  Here we consider two important (2,3)-positivity conditions, proposed by Erdahl~\cite{Erdahl, Zhao_T1}
\begin{equation}\label{T1}
T_{1} = ^{3}D + ^{3}Q \succeq 0
\end{equation}
\begin{equation}\label{T2}
T_{2} = ^{3}E + ^{3}F \succeq 0 ,
\end{equation}
where
\begin{eqnarray}\label{3pos}
^{3}D_{ijk}^{qrs} & = & \langle \psi | \hat{a}_{q}^{\dagger}\hat{a}_{r}^{\dagger}\hat{a}_{s}^{\dagger}\hat{a}_{k} \hat{a}_{j}\hat{a}_{i} | \psi \rangle \\
^{3}E_{ijk}^{qrs} & = & \langle \psi | \hat{a}_{q}^{\dagger}\hat{a}_{r}^{\dagger}\hat{a}_{s}\hat{a}_{k}^{\dagger} \hat{a}_{j}\hat{a}_{i} | \psi \rangle \\
^{3}F_{ijk}^{qrs} & = & \langle \psi | \hat{a}_{q}\hat{a}_{r}\hat{a}_{s}^{\dagger}\hat{a}_{k}\hat{a}_{j}^{\dagger} \hat{a}_{i}^{\dagger} | \psi \rangle \\
^{3}Q_{ijk}^{qrs} & = & \langle \psi | \hat{a}_{q}\hat{a}_{r}\hat{a}_{s}\hat{a}_{k}^{\dagger}\hat{a}_{j}^{\dagger} \hat{a}_{i}^{\dagger} | \psi \rangle .
\end{eqnarray}
Previous investigations indicated that the $T_{1}$ condition is less important that $T_{2}$,\cite{Zhao_T1, PhysRevA.72.052505} and therefore, it is excluded from the approximate (2,3)-positivity conditions used in this work.  The (2,2)-positivity plus the $T_{2}$ condition is denoted in this work as DQGT.

\subsection{Semidefinite Programming}

The computational implementation of the energy minimization with respect to the 2-RDM is formulated as a \textit{semidefinite program} (SDP).  A SDP is a generalization of a linear program where the objective variable is kept positive semidefinite. The program is constructed by considering the minimization of the linear energy functional in Eq.~(\ref{Efunctional}) subject to constraints
\begin{eqnarray}
\min{ {\rm Tr}(K \;X) } \\
{\rm such~that}~{\rm Tr}(A_{i} X) & = & b_{i} \label{affineC} \\
X & \succeq & 0 ,
\end{eqnarray}
where $K$ and $X$ are block matrix representations of the reduced Hamiltonian and the reduced density matrices
\begin{equation}\label{blockK2}
K = \begin{pmatrix}
 0 & 0 & 0 & 0 & 0\\
 0 & 0 & 0 & 0 & 0\\
 0 & 0 & ^{2} K & 0 & 0\\
 0 & 0 & 0 & 0 & 0\\
 0 & 0 & 0 & 0 & 0
\end{pmatrix}
\end{equation}
and
\begin{equation}\label{blockRDM}
X = \begin{pmatrix}
^{1}D & 0 & 0 &0 & 0\\
 0 & ^{1}Q & 0 & 0  & 0\\
0 & 0 &  ^{2}D &0 & 0\\
0 & 0 & 0 & ^{2}Q & 0\\
0 & 0 & 0 &0      & ^{2}G
\end{pmatrix} .
\end{equation}
The constraint matrices $A_{i}$ in Eq.~(\ref{affineC}) contain the mappings among $^{2} D$, $^{2} Q$, and $^{2} G$, the contractions to $^{1} D$ and $^{1} Q$, and the fixed trace condition.   Semidefinite programs for quantum chemical Hamiltonians have been solved with a variety of algorithms.\cite{RRSDP_Mazziotti, Zhao_T1, PhysRevA.80.032508, PhysRevLett.93.213001, PhysRevLett.106.083001} In this work we utilize the boundary-point method,\cite{PhysRevLett.106.083001,MPR08} a type of quadratic regularization method.  The floating point operations and memory scaling for the boundary-point method is $r^{6}$  for DQG (DQGT $r^{9}$) and $r^{4}$ (DQGT $r^{6}$) where $r$ is the rank of the one-electron basis set.

\section{Model}

The single-band ladder extension of the one-dimensional Hubbard model~\cite{hubbard_electron_1963, PhysRev.137.A1726, PhysRevLett.10.159} has been utilized as a minimalist model to study spin-liquid behavior\cite{PhysRevB.46.3159, Noack1996281, dagotto_surprises_1996} and high-temperature superconductors.\cite{anderson_resonating_1987, Hiroi1991230, Cava1991170, Norrestam1991864, PhysRevB.63.100402}  The ladder model is a quasi-one-dimensional system with a four-fold degenerate Fermi surface and correlations in two-dimensions.

\subsection{Hamiltonian}

The Hamiltonian of the ladder single-band model in position space is defined as follows:
\begin{eqnarray}\label{H_ladder}
\hat{H} = & ~ & t \sum_{n,\lambda,\sigma}  ( \hat{a}_{n,\lambda,\sigma}^{\dagger}\hat{a}_{n+1,\lambda,\sigma} + \hat{a}_{n+1,\lambda,\sigma}^{\dagger}\hat{a}_{n,\lambda,\sigma}) \\
& + & t_\perp \sum_{n,\sigma} (\hat{a}_{n,a,\sigma}^{\dagger}\hat{a}_{n,b,\sigma} + \hat{a}_{n,b,\sigma}^{\dagger}\hat{a}_{n,a,\sigma}) \\
& + & U\sum_{n} \hat{a}_{n,\lambda,\sigma}^{\dagger}\hat{a}_{n,\lambda,\sigma} \hat{a}_{n,\lambda,-\sigma}^{\dagger}\hat{a}_{n,\lambda,-\sigma}
\end{eqnarray}
where $t$ is a parameter controlling transport between rungs of the ladder, $t_\perp$ is a parameter controlling transport between the two sides of a ladder's rung, $U$ is a parameter controlling the one-site repulsion between electrons, the index $n$ is the rung number, the index $\lambda = a (b)$ corresponds to a ladder leg, and the index $\sigma$ indicates the spin of the electron created at rung $n$ on leg $a(b)$.   We impose periodic boundary conditions along the legs of the ladder forming a Hubbard ribbon.

\subsection{Spin and Spatial Symmetry Adaptation}

\begin{figure}[ht]
    \includegraphics[width=8.5cm]{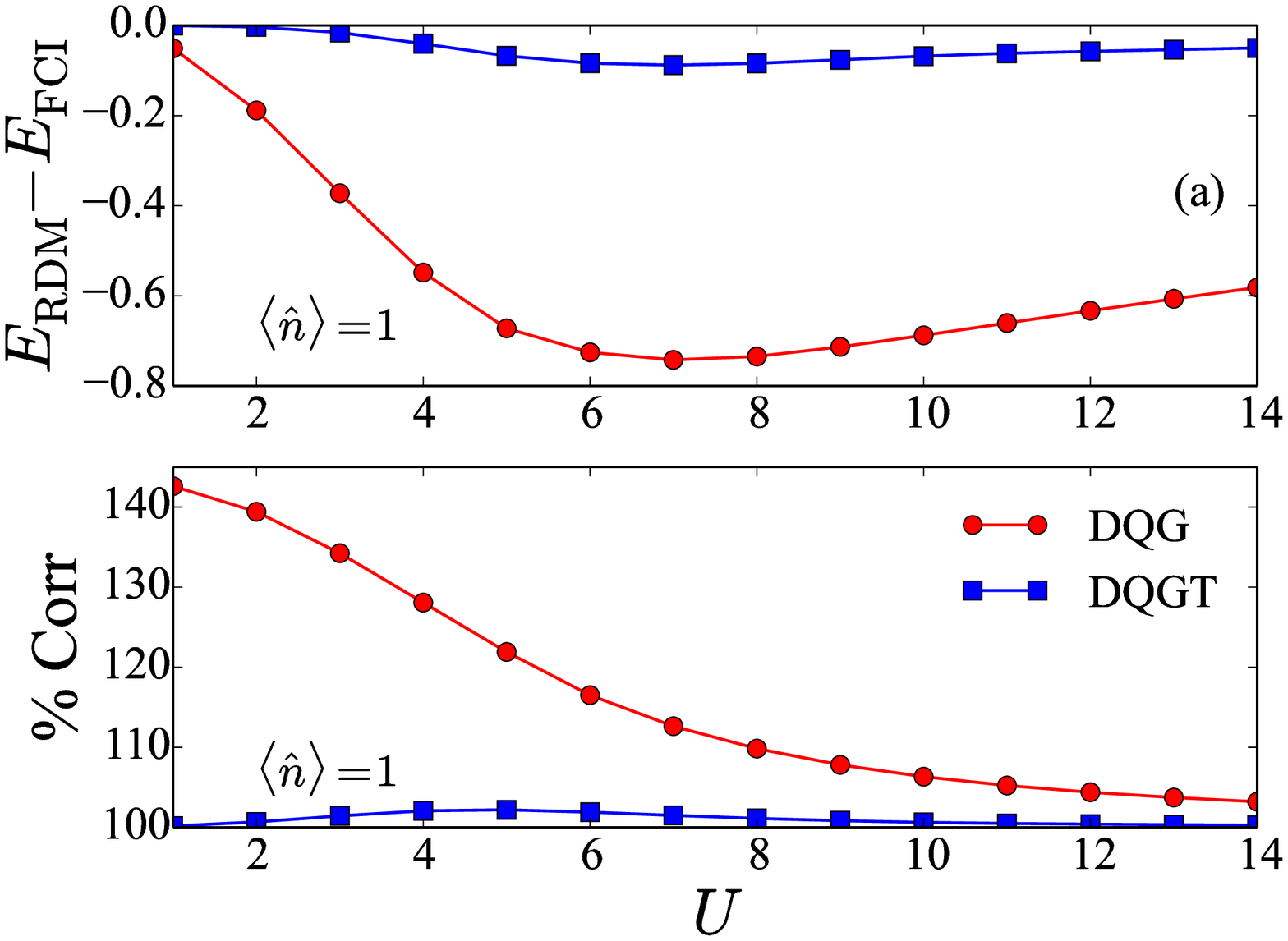}
    \includegraphics[width=8.5cm]{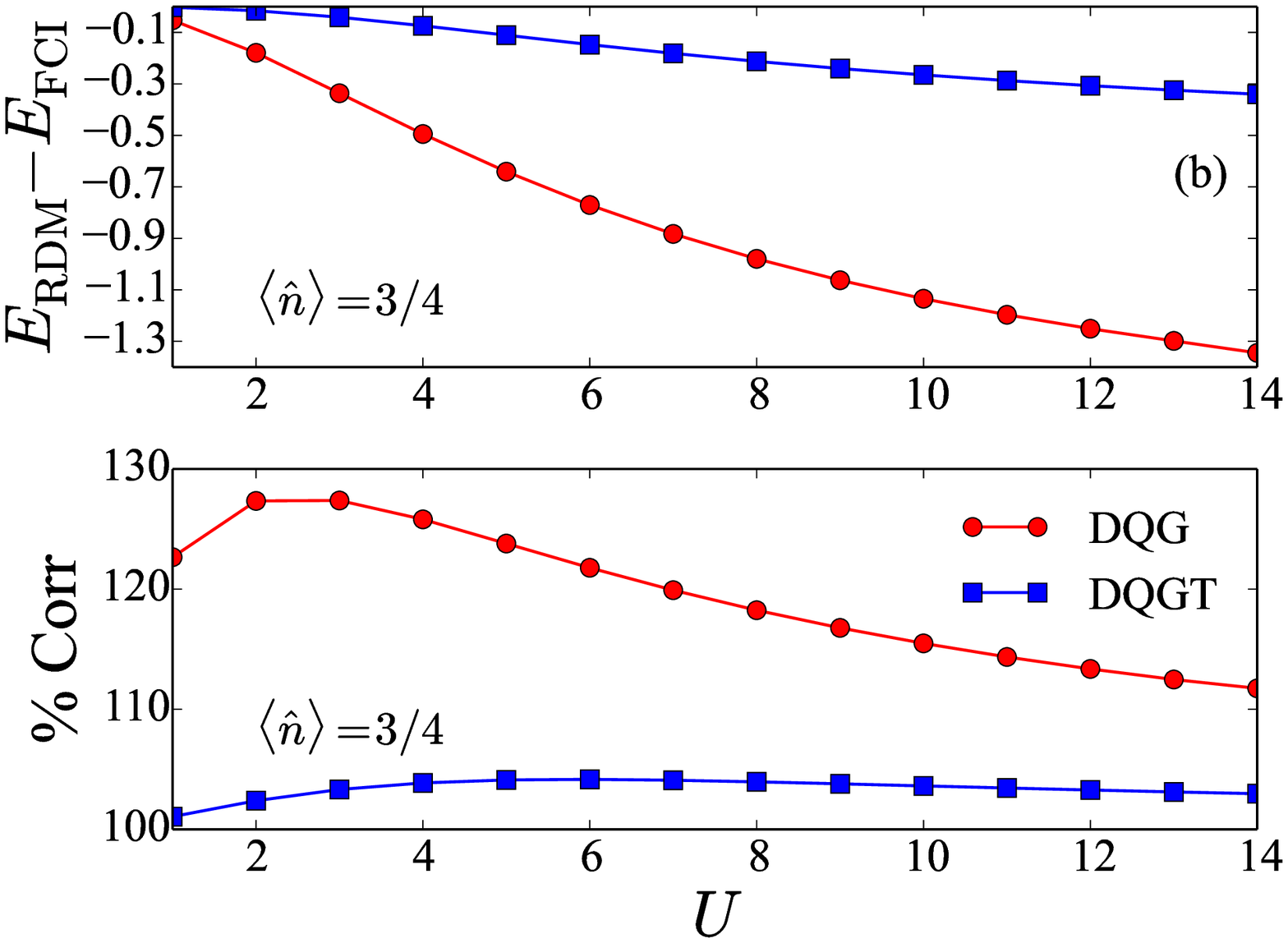}
    \caption{Absolute deviation of ground-state energy of the variational 2-RDM method with DQG and DQGT constraints and the percent correlation defined as ($E_{\mathrm{RDM}} - E_{\mathrm{HF}})/( E_{\mathrm{FCI}} - E_{\mathrm{HF}})  \times 100$ are shown for $2\times4$ lattices at (a) $\langle \hat{n} \rangle = 1$ and (b) $\langle \hat{n} \rangle = 3/4$. \label{2x4Half_filled}}
\end{figure}

One can take advantage of any spin or spatial symmetry in the Hamiltonian by symmetry adapting the metric matrices and thereby reducing the size of the 2-RDM to be optimized.\cite{PhysRevA.72.052505}  For the ladder model we transform the RDMs to bonding and antibonding spaces and then Fourier transform to take advantage of the translational symmetry.  We consider linear combination of creation and annihilation operators to form two disjoint one-electron subspaces
\begin{eqnarray}
\hat{a}_{n,\sigma}^{B} & = & \frac{1}{\sqrt{2}} \left( \hat{a}_{n,a,\sigma} + \hat{a}_{n,b,\sigma} \right) \\
\hat{a}_{n,\sigma}^{A} & = & \frac{1}{\sqrt{2}} \left( \hat{a}_{n,a,\sigma} - \hat{a}_{n,b,\sigma} \right) ,
\end{eqnarray}
where $\hat{a}_{n,\sigma}^{B}$ and $\hat{a}_{n,\sigma}^{A}$ are annihilation operators for the bonding and antibonding orbitals with spin $\sigma \in \{ \alpha, \beta \}$.  The one-body part of the Hamiltonian divides into anti-bonding ${\hat H}_{A}$ and bonding ${\hat H}_{B}$ parts:
\begin{equation}\label{BAseparation}
\hat{H} = \hat{H}_{A} + \hat{H}_{B} + \hat{H}_{\rm int}^{AB} .
\end{equation}
When expressed in the bonding and antibonding basis, the interaction term  decomposes into four two-body operators representing inter- and intra-subspace pair scattering and inter- and intra-subspace pair exchange.

Spatial symmetry is imposed in each one-electron space by the Bloch transformation
\begin{eqnarray}
\label{kdag}
^{B}\hat{a}_{n,\sigma} & = & \frac{1}{\sqrt{N_{x}}} \sum_{k_{b}} e^{-i k_{b} n} \hat{a}_{k_{b},\sigma} \\
\label{kkill}
^{A}\hat{a}_{n,\sigma} & = & \frac{1}{\sqrt{N_{x}}} \sum_{k_{a}} e^{-i k_{a} n} \hat{a}_{k_{a},\sigma} ,
\end{eqnarray}
where $\hat{a}_{k_{b},\sigma}$ annihilates an electron with momentum $k_{b}$ in the bonding band and $\hat{a}_{k_{a},\sigma}$ annihilates an electron with momentum $k_{a}$ in the antibonding band.  The Hamiltonian and 2-RDM can also be spin adapted.  As discussed in Ref.~[\onlinecite{PhysRevA.72.052505}], because the three triplet blocks are equivalent in the singlet case, each metric matrix has only two distinct spin blocks defined by the folded operators
\begin{eqnarray}
\label{spinsymm}
\hat{C}_{i,j; i\leq j}^{0,0} & = & \frac{1}{\sqrt{2}}(\hat{a}_{i,\alpha}^{\dagger}\hat{a}_{j,\beta}^{\dagger}+ \hat{a}_{j,\alpha}^{\dagger}\hat{a}_{i,\beta}^{\dagger}) \\
\hat{C}_{i,j; i<j}^{1,0} & = & \frac{1}{\sqrt{2}}(\hat{a}_{i,\alpha}^{\dagger}\hat{a}_{j,\beta}^{\dagger}+ \hat{a}_{j,\alpha}^{\dagger}\hat{a}_{i,\beta}^{\dagger}) .
\end{eqnarray}
These new $\hat{C}$ operators are substituted into Eq.~(\ref{metricOverlap}) generating symmetric and antisymmetric parts of $^{2} D$ and $^{2} Q$.  Spin-symmetry adaptation of $^{2} G$ and $T_{2}$ can be achieved by the same methodology.\cite{PhysRevA.72.032510}  The size of the 2-RDM can be further reduced by additional symmetries,\cite{Verstichel201312, verstichel_v2dm_2013, PhysRevLett.67.3848} but they have not been exploited in the present calculations.

\section{Results}

In section~\ref{sec:Energy2x4} we compare the ground-state energies of the $2\times4$ ladder system at and below half filling from the variational 2-RDM method with those from full configuration interaction (FCI).  Section \ref{sec:Corrs} contains the analysis of the $2\times10$ and $1\times10$ lattices through the $\alpha,\beta$-two-point pair correlation function, a measure of one-particle effective hopping, one-electron natural occupation numbers, and the squared Frobenius norm of the cumulant (connected) part of the 2-RDM.  Results from 2-RDM calculations with DQG and DQGT conditions are compared.

\subsection{Energies of Hubbard Ladder}

\label{sec:Energy2x4}

\begin{figure}[ht]
  \includegraphics[width=8.5cm]{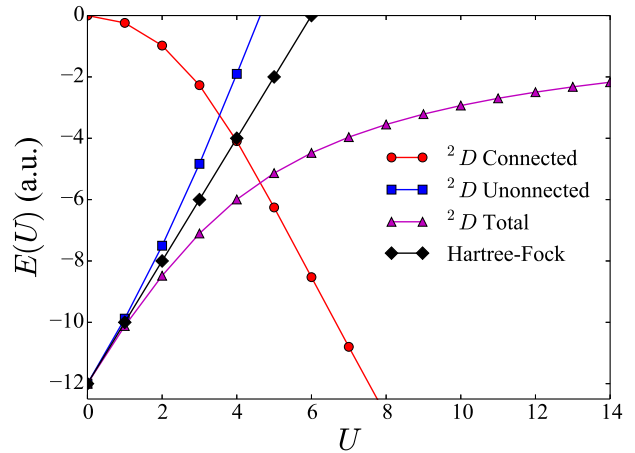}
  \caption{For the $2\times4$ lattice at $\langle \hat{n} \rangle = 1$ the connected, unconnected, and total energies from the variational 2-RDM method with DQGT constraints as well as the Hartree-Fock total energies are shown. \label{E_2x10}}
\end{figure}

We report the errors in the ground-state energies from the variational 2-RDM method for the $2\times4$ ladder Hubbard model for a range of interaction strengths where $t = t_{\perp} = 1$~a.u.  Comparisons are made to the ground-state energies from full configuration interaction (FCI).  The FCI calculation determines the ground-state energies by computing the lowest eigenvalue of the $N$-electron Hamiltonian matrix in the basis of all possible Slater determinants.  In the FCI calculation the spin orbitals are products of a spin function ($\alpha$ or $\beta$) and a ``spatial'' orbital which can be defined either in the position representation or in the momentum representation.  All 2-RDM alculations were optimized until the primal feasibility norm was below $1.0 \times 10^{-5}$ and the primal-dual gap was below $1.0 \times 10^{-4}$.  In Fig.~\ref{2x4Half_filled} the error in the ground-state energy and the percentage of the correlation energy recovered from the 2-RDM method with DQG and DQGT conditions is reported for (a) $\langle \hat{n} \rangle = 1$ and (b) $\langle \hat{n} \rangle = 3/4$ fillings.  In the case of half-filling the DQG and the DQGT energies deviate at most from those from FCI by -0.74~a.u. and -0.087~a.u., respectively.  The maximum deviations occur in the intermediate interaction regime $U \in [4,8]$ where there is a large degree of competition between delocalization and localization.  For the $\langle \hat{n} \rangle = 3/4$ filling the DQG and DQGT conditions result in a larger absolute error than all corresponding values at half filling, which is consistent with previous observations in the literature\cite{verstichel_variational_2012} that correlated systems with an imbalance between the number of particles and holes require more stringent $N$-representability constraints.

The 2-RDM can be expressed as the wedge product of 1-RDMs (unconnected) plus a cumulant (connected) part\cite{M98b,M98c} denoted as $^{2}\Delta$
\begin{equation}\label{E_conn_unconn_total}
^{2}D = \,^{1}D \wedge \, ^{1}D + \,^{2}\Delta ,
\end{equation}
where the $\wedge$ is the Grassmann wedge product~\cite{Sleb70,CA80,M98a}.  The unconnected term captures the statistically independent part of the electron pair probability.   The energies from the unconnected and connected components\cite{hammond_variational_2005} are
\begin{eqnarray}
\label{E1}
E_{1} & = & {\rm Tr}[^{2}K\;(^{1}D \wedge\,^{1}D)] \\
\label{E2}
E_{2} & = & {\rm Tr}[ ^{2}K\; ^{2}\Delta] .
\end{eqnarray}
These energies as well as the Hartree-Fock mean-field energy are plotted in Fig.~\ref{E_2x10}.  The Hartree-Fock energy grows linearly as $U$ is increased which is closely mirrored by the unconnected piece.  Consequently, all the correlated information of the 2-RDM that results in localization is contained in its connected part.

\subsection{One- and Two-particle Correlations}\label{sec:Corrs}

\begin{figure}
    \includegraphics[width=8.5cm]{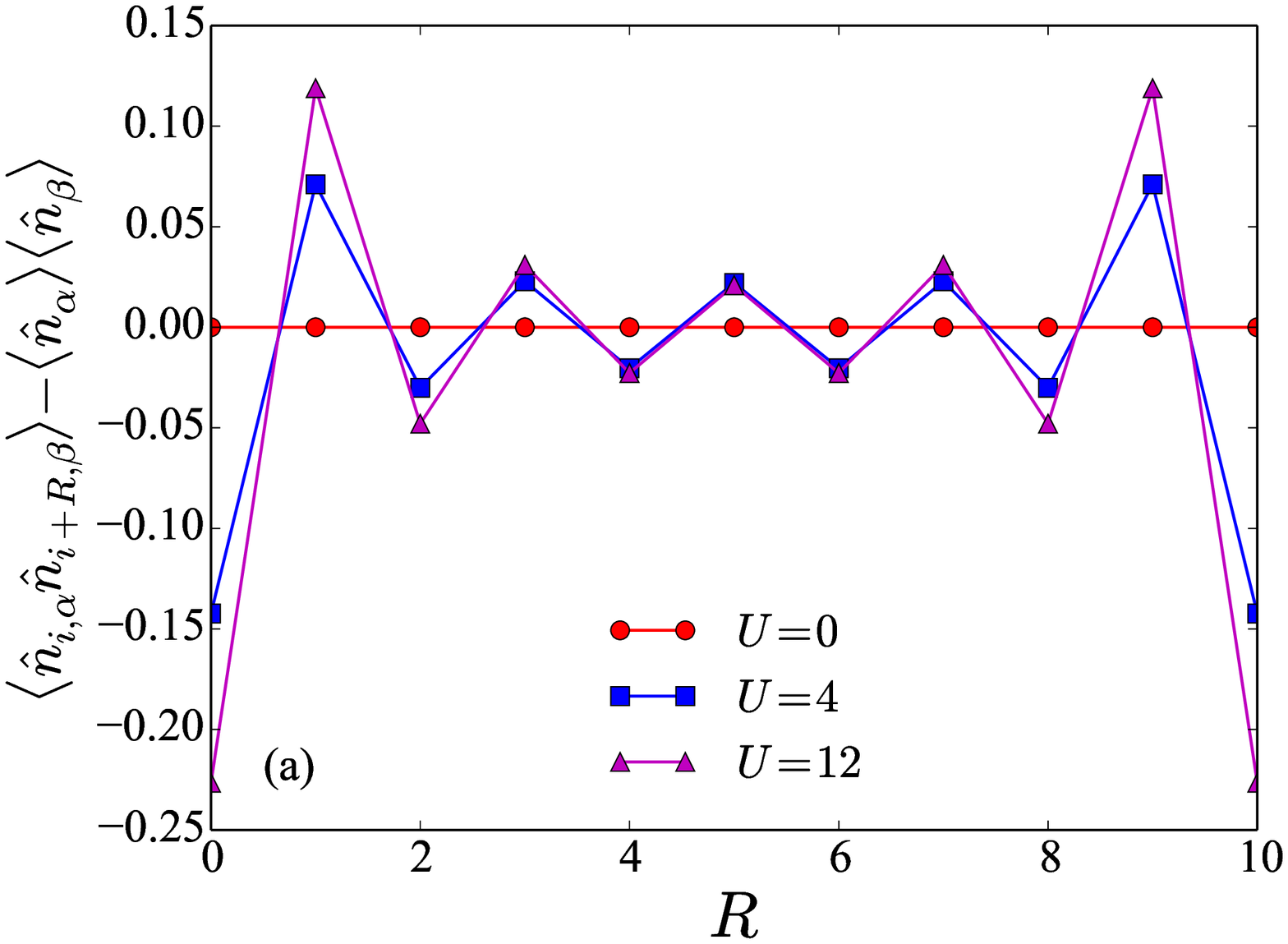}
    \includegraphics[width=8.5cm]{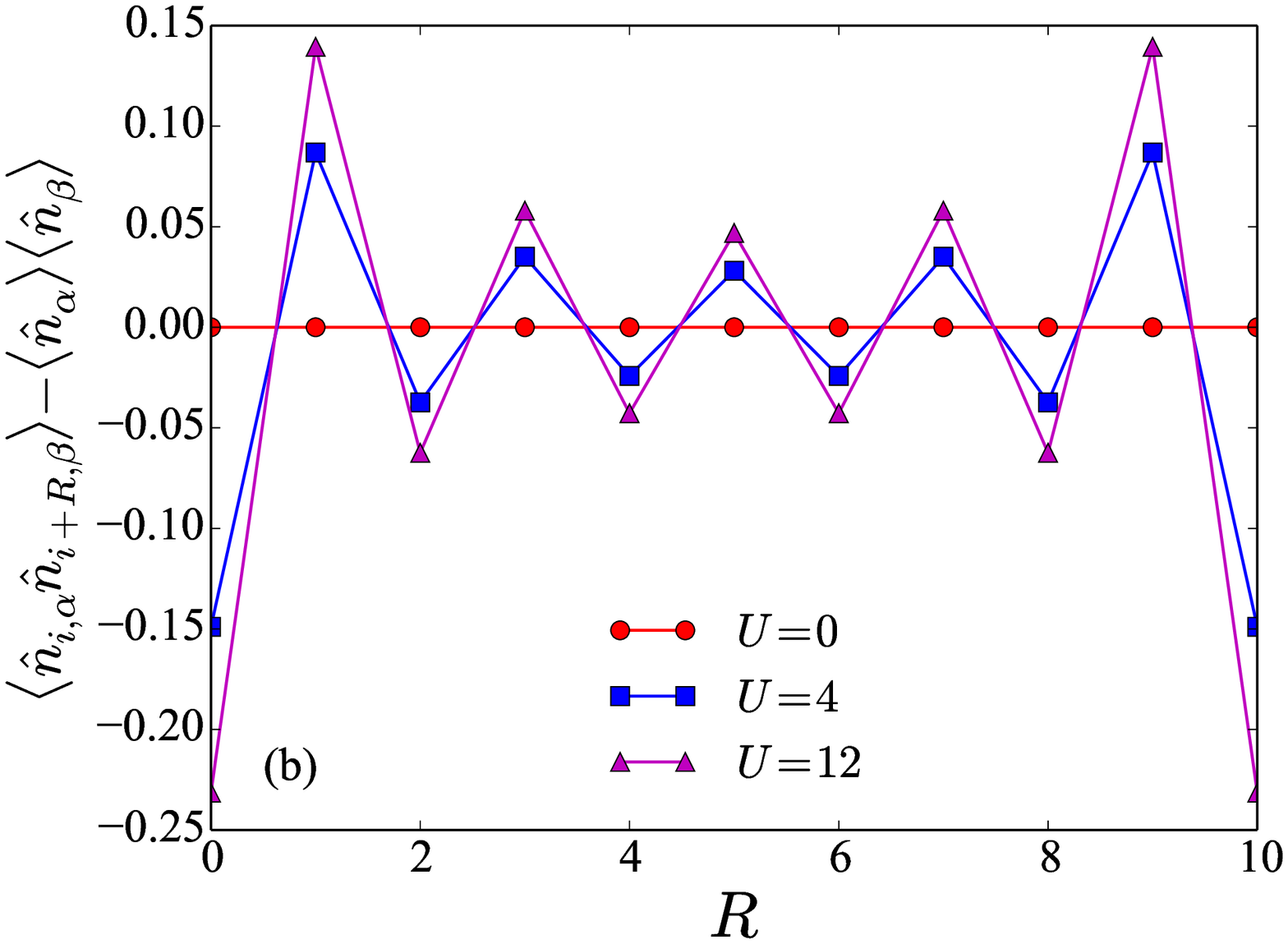}
    \caption{The $\alpha,\beta$-two-point pair correlation functions of the (a) $2\times10$ and (b) $1\times10$ lattices at half filling are computed as a function of $R$ from the variational 2-RDM method with DQGT where $R$ is the distance along each Hubbard strand between the pairs.  Because of the periodic boundary conditions, the correlation function values are unique until the lattice inversion center at $R=5$.\label{fig:DQGT2x10M10}}
\end{figure}

The $\alpha,\beta$-two-point pair correlation function for the $2\times10$ and $1\times10$ lattices is examined at half filling to explore the variational 2-RDM method with DQG and DQGT conditions.  We can express the two point spin-up, spin-down correlation as
\begin{equation}\label{canonical_spincorr}
\langle \hat{n}_{i\alpha}\hat{n}_{i+R\beta} \rangle - \langle \hat{n}_{\alpha} \rangle\langle \hat{n}_{\beta} \rangle ,
\end{equation}
where $\langle \hat{n}_{\alpha} \rangle$ is the total density of $\alpha$-electrons in the system
\begin{equation}
\langle \hat{n}_{\alpha} \rangle = N_{\alpha}/(2 N_{L}) , \langle \hat{n}_{\beta} \rangle = N_{\alpha}/(2 N_{L}) .
\end{equation}
This correlation function is an extension of the local double-occupancy, $\langle \hat{n}_{i,\alpha}\hat{n}_{i,\beta} \rangle$, which has been used to examine the Mott transition of the Hubbard model defined on various lattices and temperatures.~\cite{PhysRevLett.97.066401, PhysRevLett.108.246402, PhysRevA.82.043625, PhysRevB.82.245102}

\begin{figure*}
    \includegraphics[width=8.5cm]{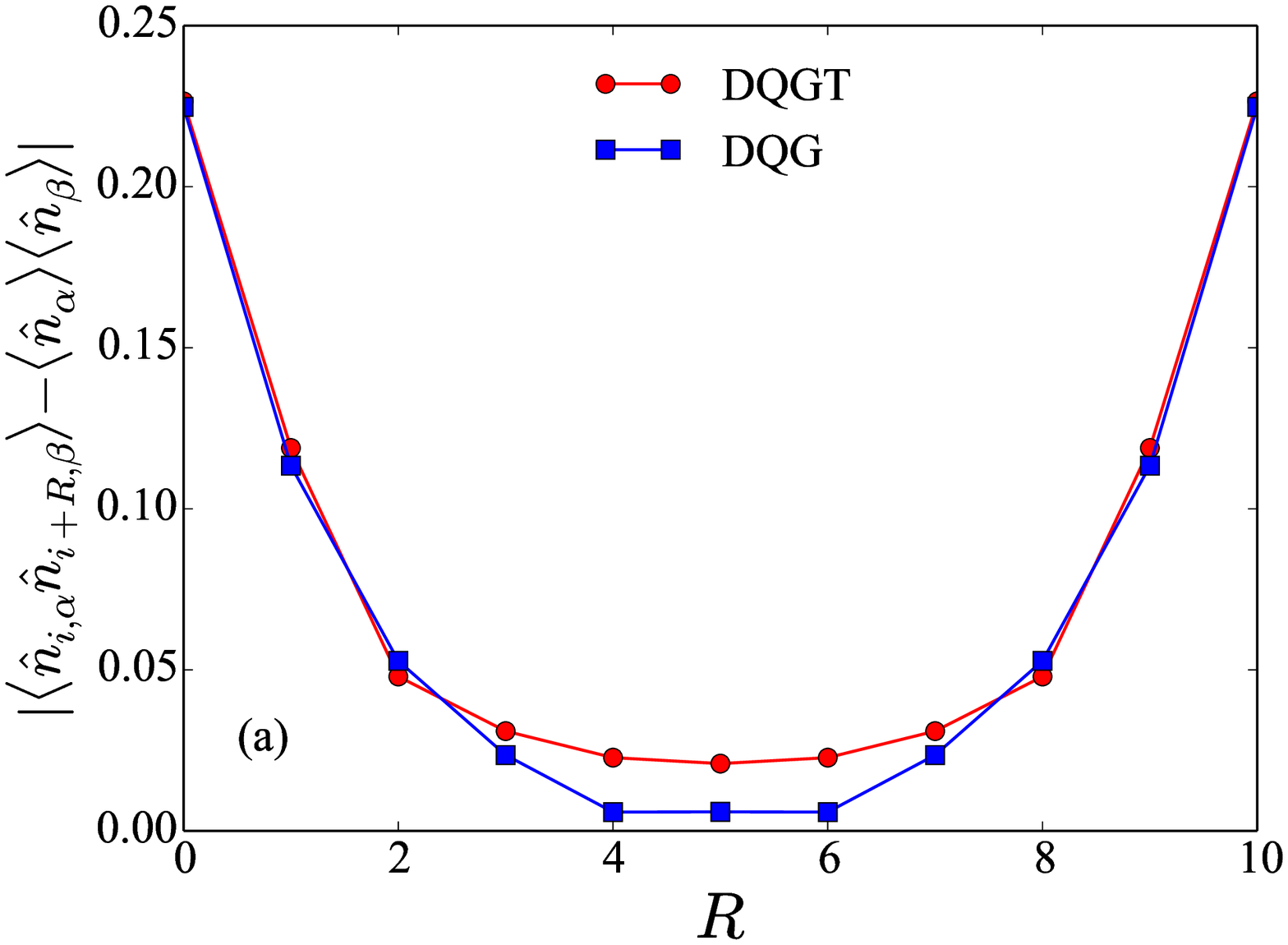}
    \includegraphics[width=8.5cm]{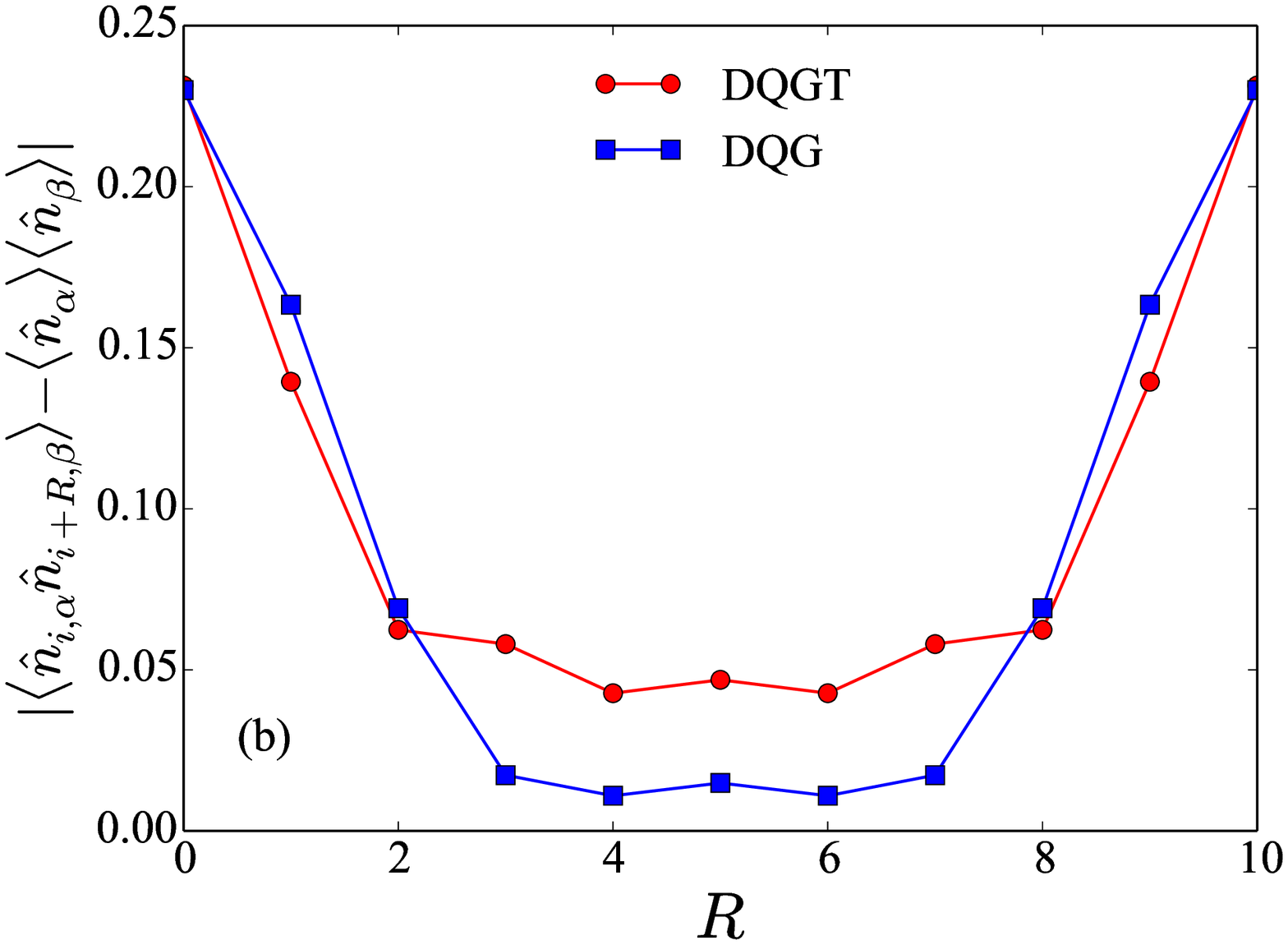}
    \caption{The $\alpha,\beta$-two-point pair correlation functions calculated with DQG and DQGT constraints are given for the (a) ladder $2\times10$ and (b) linear $1\times10$ lattices at half filling $\langle \hat{n} \rangle = 1$.\label{DQGTvsDQG2x10M10} }
\end{figure*}

The pair-correlation function probes the expectation of an antiferromagnetic pair $R$-sites away from each other in the long direction of the lattice.   In the $U = 0$ limit where the $\alpha$ and $\beta$ electrons are delocalized across the lattice the correlation function is zero for all values of $R$.  As $U$ is increased, antiferromagnetic pairing becomes less favorable and order is induced.  In Figs.~\ref{fig:DQGT2x10M10}a and~\ref{fig:DQGT2x10M10}b we plot the two-particle correlation function as a function of $R$ for representative $U$ values.  The correlation function decays across the lattice until the inversion center is reached.  An exponential fitting of the absolute value of the pair correlation function, not shown, indicates that DQGT predicts a similar decay on the one-dimensional and  quasi-one-dimensional lattices.  In Fig.~\ref{DQGTvsDQG2x10M10} we plot the absolute value of the pair-correlation function for $U=12$ from DQG and DQGT to study the differences generated in the (a) ladder and (b) linear lattices from approximate $N$-representability conditions.  In the ladder case the results from DQG and DQGT are in fairly close agreement while in the linear case DQG deviates significantly from DQGT.

\begin{figure*}
  \includegraphics[width=8.5cm]{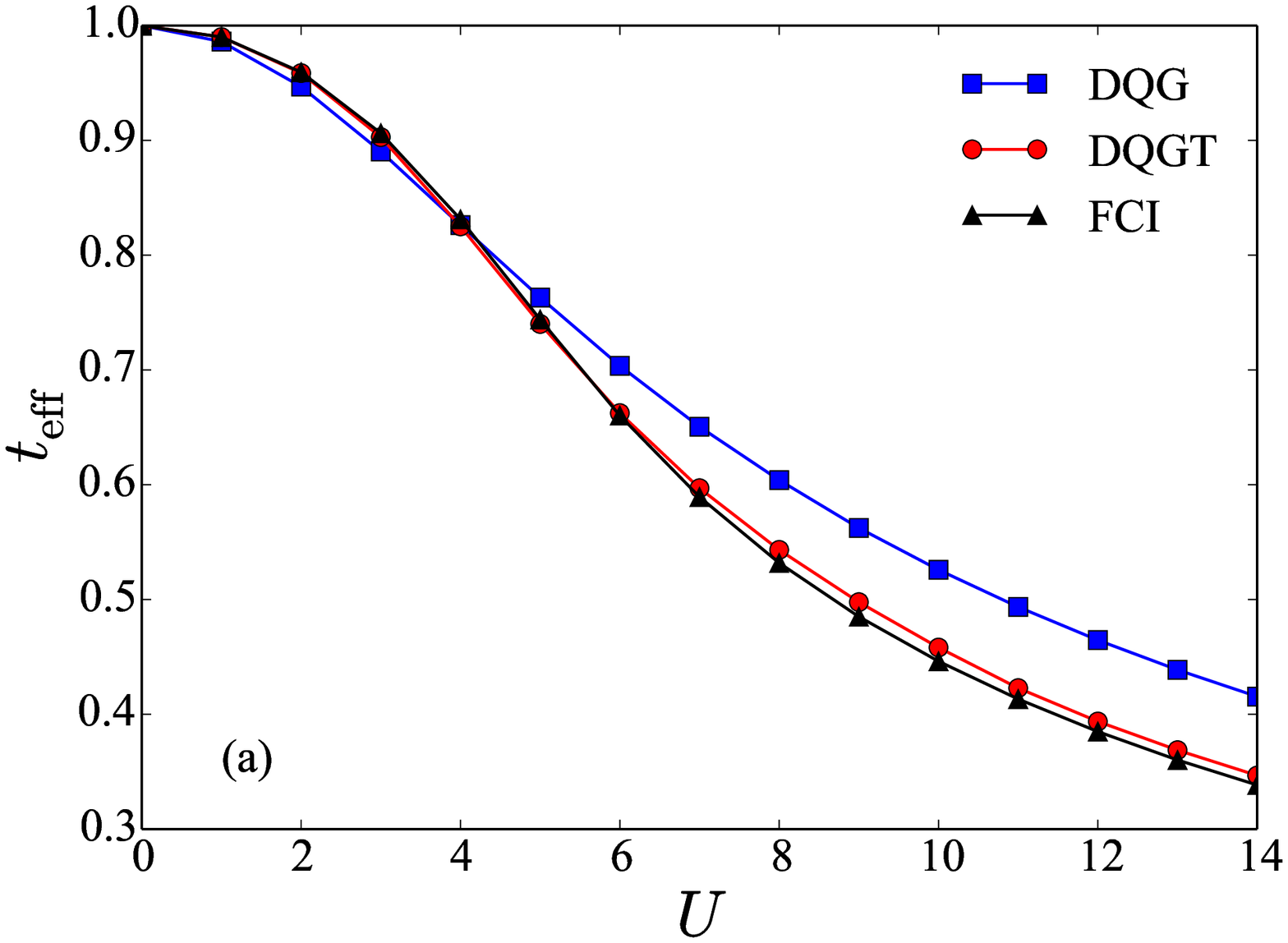}
  \includegraphics[width=8.5cm]{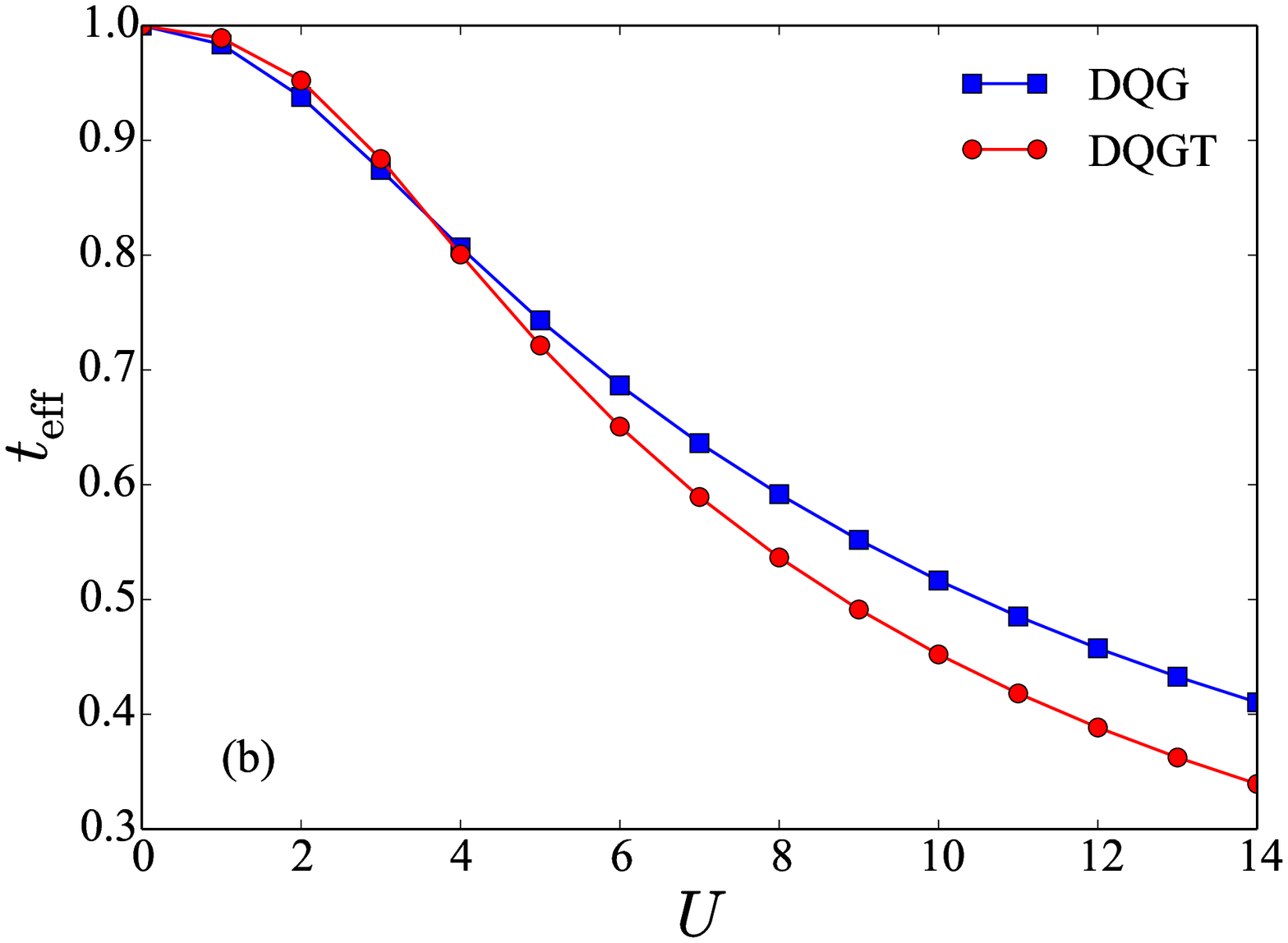}
  \includegraphics[width=8.5cm]{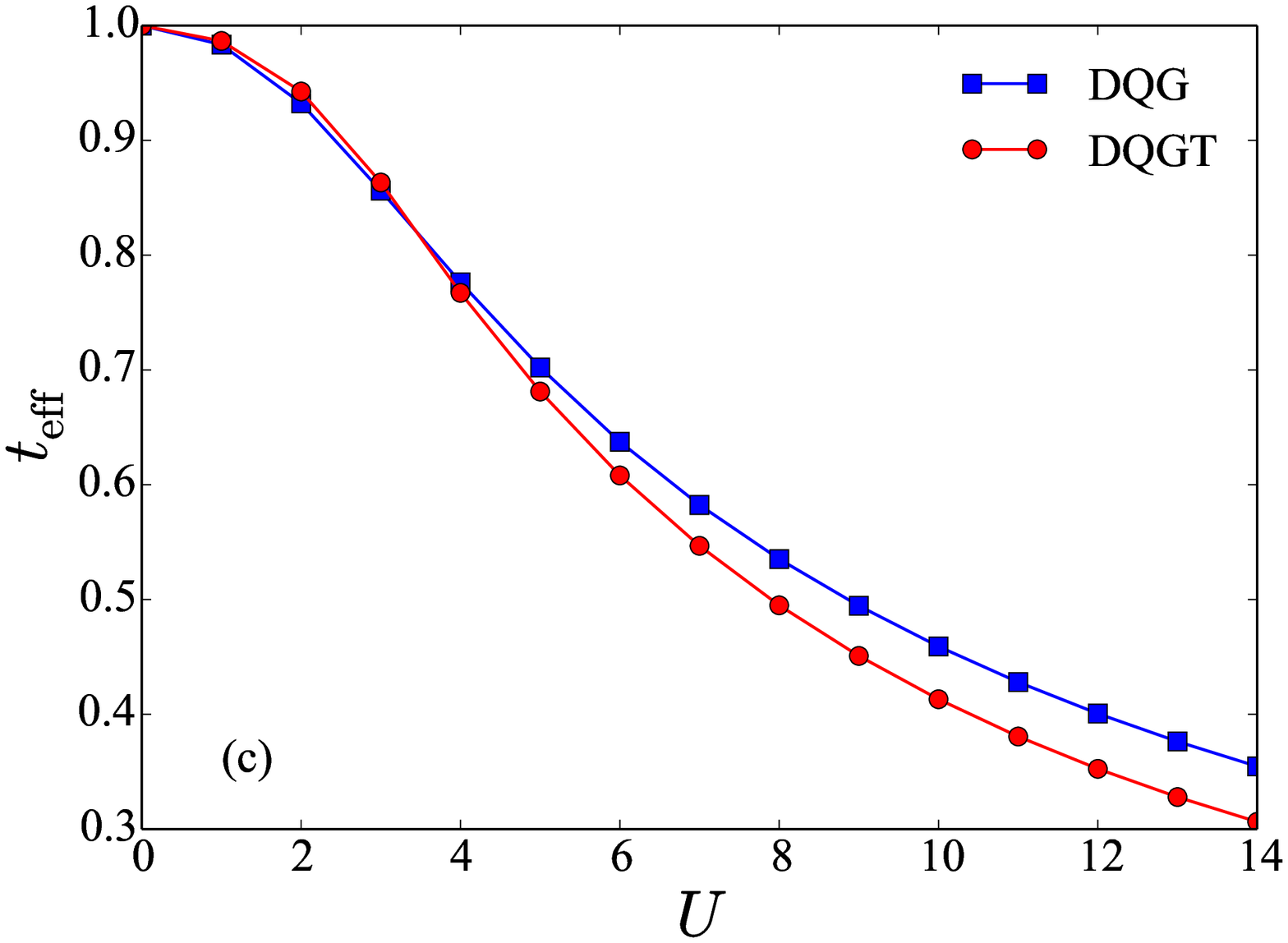}
  \includegraphics[width=8.5cm]{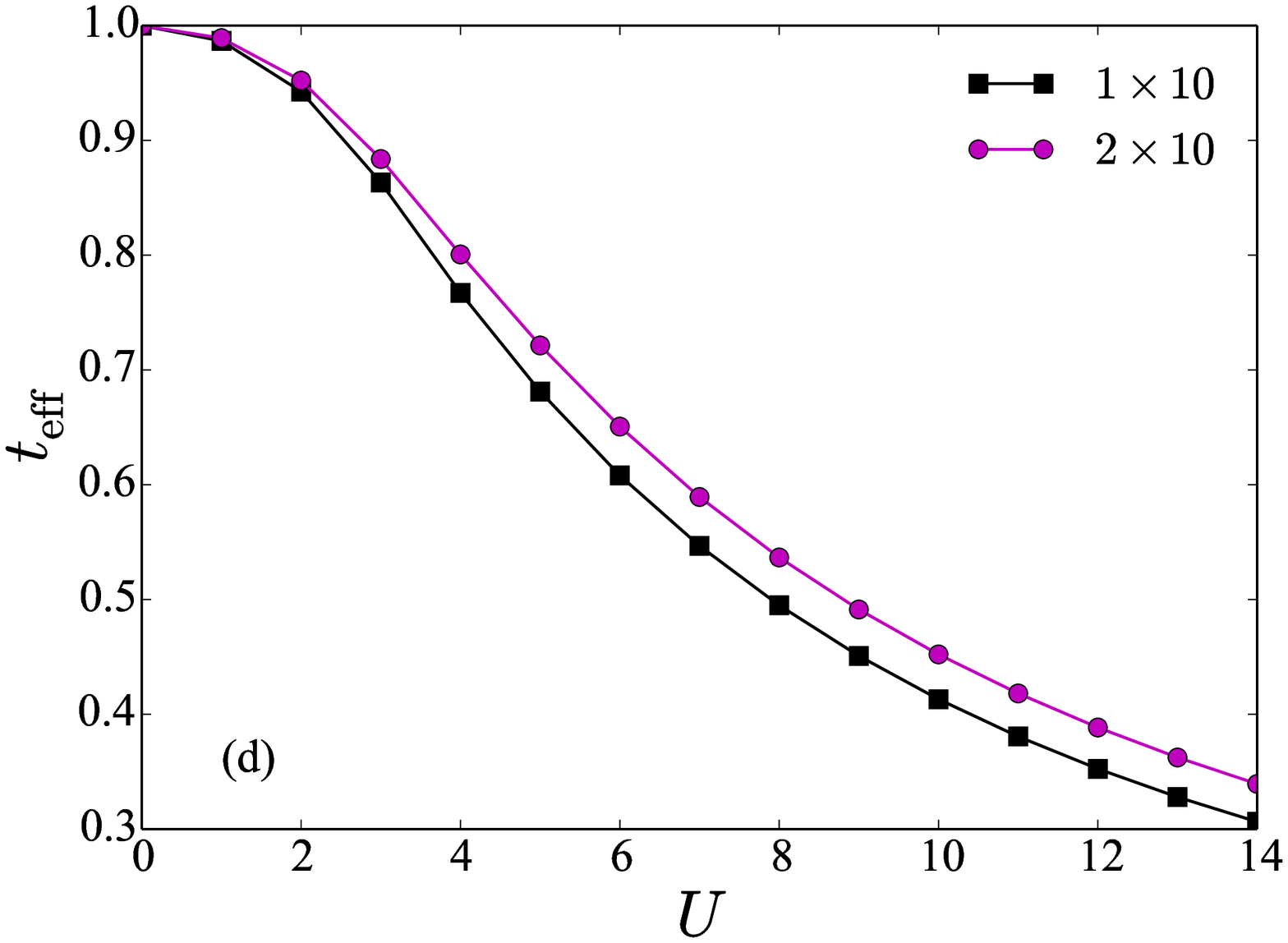}
  \caption{The effective hopping $t_{\mathrm{eff}}$ for the $2\times4$ lattice is calculated with DQG, DQGT, and FCI.  In (a) we plot $t_{\mathrm{eff}}$ for the $2\times4$ lattice calculated with DQG and DQGT compared against $t_{\mathrm{eff}}$ from FCI. We compare $t_{\mathrm{eff}}$ from DQG and DQGT for the (b) $2\times10$ and (c) $1\times10$ lattices.   Both the linear and ladder Hubbard models exhibit the same underestimation of $t_{\mathrm{eff}}$ at low $U$ and overestimation of $t_{\mathrm{eff}}$ at high $U$.  Part (d) compares $t_{\mathrm{eff}}$ calculated from DQGT for the linear and ladder models. The increase in lattice dimension facilitates transport down the chain. \label{teff} }
\end{figure*}

We can use the effective hopping as a one-particle correlation function.   Defined in Eq.~(\ref{teffteff}), the effective hopping is the likelihood of transport normalized by the non-interacting limit of the model
\begin{equation}\label{teffteff}
t_{\mathrm{eff}} = \frac{\langle \hat{a}_{i\sigma}^{\dagger}\hat{a}_{j\sigma} + \hat{a}_{j\sigma}^{\dagger}\hat{a}_{i\sigma} \rangle _{U} }{ \langle \hat{a}_{i\sigma}^{\dagger}\hat{a}_{j\sigma} + \hat{a}_{j\sigma}^{\dagger}\hat{a}_{i\sigma} \rangle_{U = 0}} .
\end{equation}
For low $U$, where the kinetic energy has the largest energy contribution, approximate $N$-representability conditions underestimate the effective hopping (kinetic energy), and at high $U$, where the repulsive interaction term dominates, approximate $N$-representability conditions underestimate the localization, which causes the effective hopping to be overestimated.  In Fig.~\ref{teff}a we plot $t_{\mathrm{eff}}$ for the $2\times4$ lattice calculated with DQG and DQGT compared against $t_{\mathrm{eff}}$ from FCI.  As expected DQG yields a lower $t_{\mathrm{eff}}$ than FCI before $U = 4$ and a higher $t_{\mathrm{eff}}$ than FCI after $U = 4$.  The effective hopping from DQGT shows a similar trend while exhibiting a much smaller deviation from the FCI curve, which reinforces the accuracy of the DQGT conditions for lattices with strongly correlated electrons.  We compare $t_{\mathrm{eff}}$ from DQG and DQGT for the (b) $2\times10$ and (c) $1\times10$ lattices.   Both the linear and ladder Hubbard models exhibit the same underestimation of $t_{\mathrm{eff}}$ at low $U$ and overestimation of $t_{\mathrm{eff}}$ at high $U$.  Figure~\ref{teff}d compares $t_{\mathrm{eff}}$ calculated from DQGT for the linear and ladder models.  The increase in lattice dimension facilitates transport down the chain.

\subsection{Natural Occupation Numbers and Entanglement}

\begin{table}
    \caption{Natural-orbital occupation numbers of the $\alpha$-spin block of the 1-RDM in the quasi-momentum basis.\label{RDM1Occs}}
    \begin{ruledtabular}
    \begin{tabular}{lccccccc}
       Lattice  & & Conditions & $U$       & $N_{\alpha}-1$    & $N_{\alpha}$      & $N_{\alpha}+1$      & $N_{\alpha}+2$    \\
    \hline
    $1\times10$ & & DQG       & 4         & 0.9022        &	0.8064  &	0.1936  &	0.0978      \\
                & &           & 8         & 0.7828        &	0.6566  &	0.3434  &	0.2172      \\
                & &           & 12        & 0.7114        &   0.6014  &   0.3986  &   0.2886    \\
                & &    DQGT   & 4         & 0.8993        &	0.7895  &	0.2104  &	0.1008      \\
                & &           & 8         & 0.7634        &   0.6291  &   0.3710  &   0.2366        \\
                & &           & 12        & 0.6870        &	0.5812  &	0.4188  &	0.3130      \\
    $2\times10$ & & DQG       & 4         & 0.8081        &	0.7667  &	0.2333  &	0.1919      \\
                & &           & 8         & 0.6629        &	0.6280  &	0.3720  &	0.3371      \\
                & &           & 12        & 0.6067        &   0.5827  &   0.4173  &   0.3933      \\
                & & DQGT      & 4         & 0.7863        &	0.6897  &	0.3103  &	0.2138      \\
                & &           & 8         & 0.6338        &	0.5710  &	0.4290  &	0.3662      \\
                & &           & 12        & 0.5811        &	0.5462  &	0.4538  &	0.4189
    \end{tabular}
    \end{ruledtabular}
\end{table}

We examine the one-electron occupation numbers of the natural orbitals around the Fermi surface.  The natural orbitals are the eigenfunctions of the 1-RDM. Select occupation numbers of the $\alpha$-spin block of the 1-RDM in the quasi-momentum basis are provided in Table~\ref{RDM1Occs}.   For both lattices the $N_{\alpha}-1$ and $N_{\alpha}$ occupation numbers calculated with DQG are larger than the occupation numbers with DQGT, and the $N_{\alpha}+1$ and $N_{\alpha}+2$ occupation numbers calculated with DQG are smaller than the occupation numbers with DQGT.  The difference between DQG and DQGT occupation numbers is greater for the $2\times10$ lattice than $1\times10$ lattice for all $U$ values.  Furthermore, the higher degree of multireference character of the occupation numbers when calculated with DQGT constraints indicates that polyradical character induced by a transition to the Mott-insulating state is better captured by DQGT conditions.

\begin{figure*}
    \includegraphics[width=8.5cm]{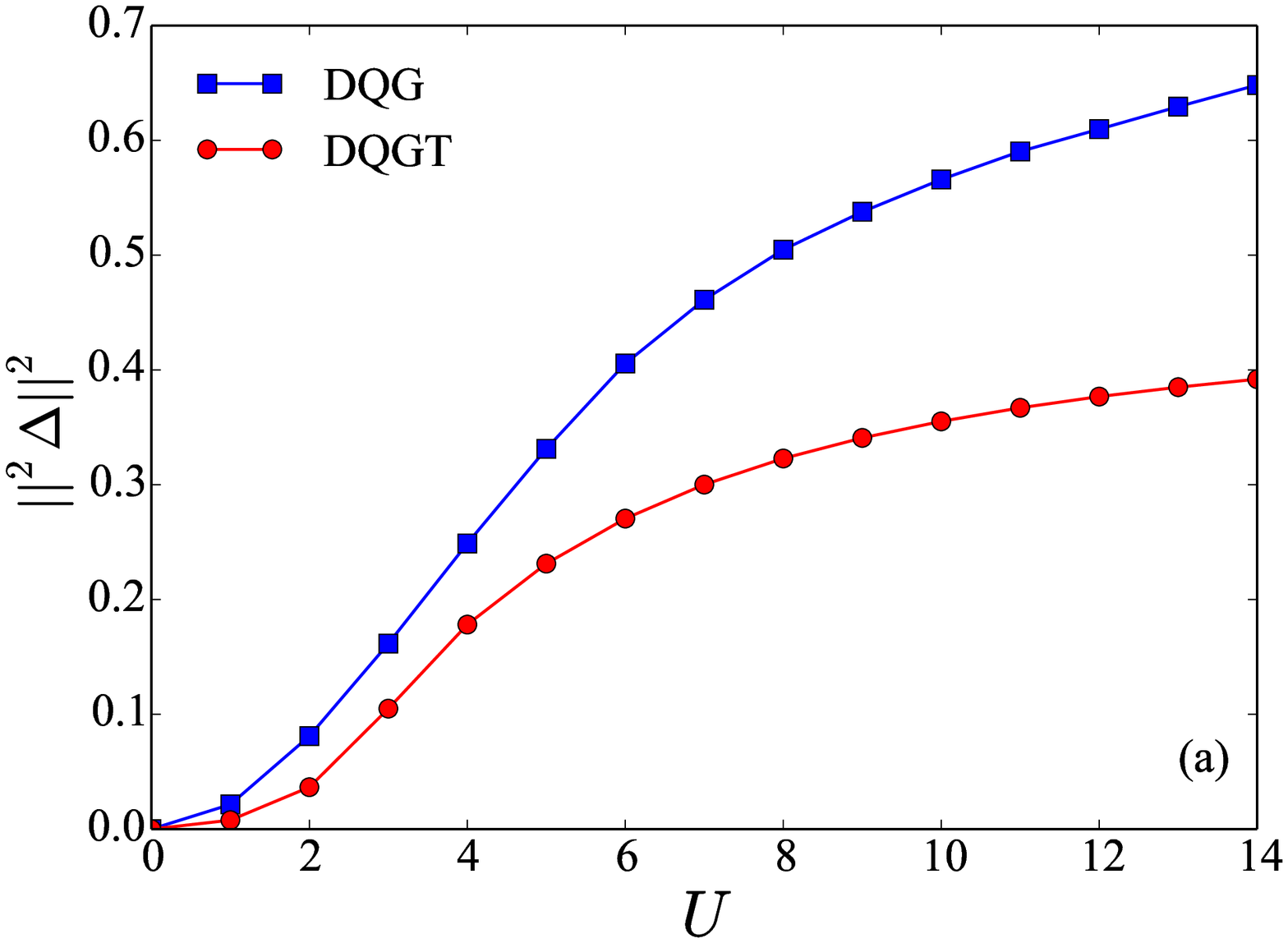}
    \includegraphics[width=8.5cm]{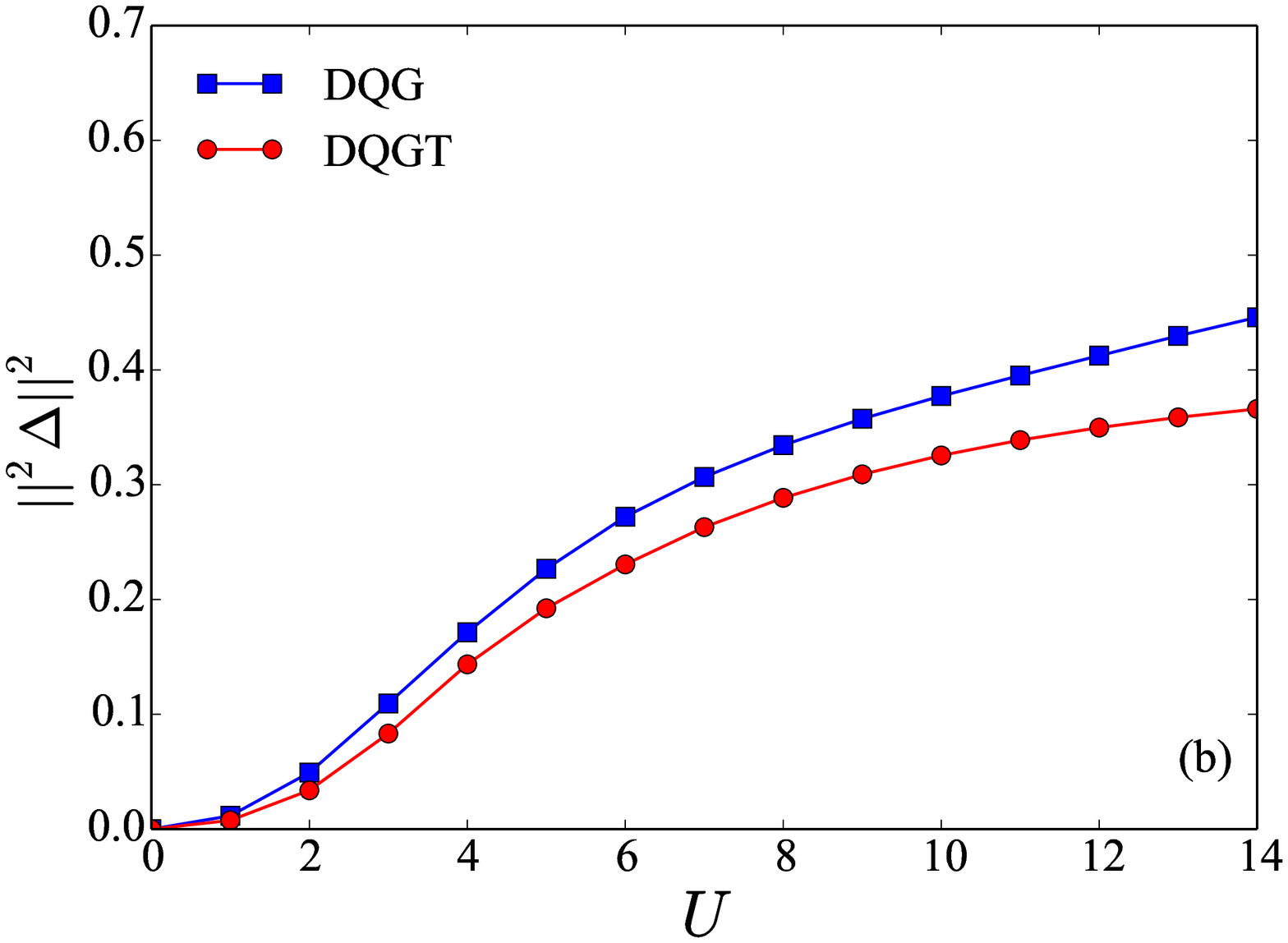}
    \caption{Frobenius norm squared of the cumulant part of the 2-RDM is shown for the (a) $2\times10$ and (b) $1\times10$ lattices where the 2-RDMs are computed with variational 2-RDM calculations with DQG and DQGT conditions. \label{FNS_1x10} }
\end{figure*}

We measure correlation explicitly by calculating the squared Frobenius norm of the cumulant portion of the 2-RDM $^{2}\Delta$
\begin{equation}\label{FNSeq}
||^{2}\Delta||^{2} = {\rm Tr}(^{2}\Delta^{\dagger} \; ^{2}\Delta)
\end{equation}
as a function of system size and coupling.  The squared Frobenius norm of the cumulant\cite{TJ_JCP1252006} is size-extensive and contains spin entanglement information not captured by the correlation energy or von Neumann entropy.  Using this metric, we compare the degree of correlation in each ladder system.  In Fig.~\ref{FNS_1x10} we plot the norm-per-lattice site for the (a) $2\times10$ and (b) $1\times10$ lattices at half filling.  For the $2\times10$ lattice DQG predicts a significantly larger norm at high $U$.  In the case of the $1\times10$ lattice DQG and DQGT agree on the amount of correlation in the lattice.

\section{Conclusion}

Calculation of ground-state properties of strongly correlated model systems is highly important for understanding a plethora of condensed phase $N$-body physics.  One of the main limitations of wave function methodologies is the exponential scaling of the Hilbert space with system size.  In exchange for the exponentially scaling $N$-particle wave function we have reviewed how to compute the ground-state energy with respect to the polynomial scaling 2-RDM.  The variational 2-RDM has some key benefits: ({\em i}) it provides a lower bound to wave function methods, ({\em ii}) can be numerically implemented as a semidefinite program which is solved with a polynomial scaling algorithm, and ({\em iii}) and gives easy access to pair correlation functions important for characterizing condensed-matter systems.

We have demonstrated that the variational 2-RDM method with moderate $N$-representability constraints can be used to calculate the ground state energies of ladder Hubbard models accurately.  In keeping with recent results for $4\times4$ and $6\times6$ two-dimensional Hubbard models, we observe that partial (2,3)-positivity (DQGT) conditions are effective at capturing strong electron correlation effects in both one- and quasi-one-dimensional lattices for both half filling and less-than-half filling.  We have found that certain correlation functions can be accurately predicted with (2,2)-positivity (DQG) conditions.  Furthermore, 2-RDM methods offer a way to analyze the correlation per site in a lattice model with a size-extensive metric and give direct access to occupation numbers.  The 2-RDM methods complement recently developed wave-function-based methods~\cite{U12}, and they may be useful in the context of approximate embedding calculations.\cite{Scus13}  In general, the variational RDM method offers a new approach to studying lattice models of varying topology and filling.

\begin{acknowledgments}

D.A.M. gratefully acknowledges the NSF, the ARO, and the Keck Foundation for their generous support.  The authors express their gratitude to Dr. Jonathan J. Foley for assistance with the configuration interaction calculations.

\end{acknowledgments}

\bibliography{biblo}

\end{document}